\documentclass[12pt]{iopart}

\usepackage{graphicx}
\usepackage{xcolor}

\begin{document}

\title{Finding Proper Time Intervals for Dynamic Network Extraction}

\author{G\"{u}nce Keziban Orman\textsuperscript{*}, Nadir Türe, Selim Balcisoy, Hasan Alp Boz}

\address{Computer Engineering Department, Galatasaray University, Ortak\"{o}y, Istanbul, Turkey, Computer Engineering Department, Sabancı University Tuzla, Istanbul }
\ead{\textsuperscript{*}korman@gsu.edu.tr, nture92@gmail.com, balcisoy@sabanciuniv.edu, bozhasan@sabanciuniv.edu}

\begin{abstract}
Extracting a proper dynamic network for modelling a time-dependent complex system is an important issue. Building a correct model is related to finding out critical time points where a system exhibits considerable change. In this work, we propose to measure network similarity to detect proper time intervals. We develop three similarity metrics, node, link, and neighborhood similarities, for any consecutive snapshots of a dynamic network. Rather than a label or a user-defined threshold, we use statistically expected values of proposed similarities under a null-model to state whether the system changes critically. We experimented on two different data sets with different temporal dynamics: The Wi-Fi access points logs of a university campus and Enron emails. Results show that, first, proposed similarities reflect similar signal trends with network topological properties with less noisy signals, and their scores are scale invariant. Second, proposed similarities generate better signals than adjacency correlation with optimal noise and diversity. Third, using statistically expected values allows us to find different time intervals for a system, leading to the extraction of non-redundant snapshots for dynamic network modelling.
\end{abstract}

%

%
%
%

\section{Introduction}\label{Introduction}
Dynamic network representation has attracted the attention of many scientists from different domains \cite{Spiliopoulou2011, Blonder2012, Holme2012, Aggarwal2014,Leskovec2007,Kempe:2000,Rossi2013,BergerWolf2006}. They allow us to consider not only the system objects and interactions but also their changes. Several different representations such as event lists, link streams, or event-based sequential graphs are proposed \cite{Holme2012}. The most common one is time-based sequential graph modelling in which each member of the network sequence, a.k.a \textit{snapshot}, contains interactions between the observed system actors for a given \textit{time interval}. In this work, we focus on proper dynamic network extraction for time-based sequential graph modelling.

\textit{Proper dynamic network extraction} is a challenging issue because it is related to making decisions on the model components such as nodes, links, or time intervals when extracting snapshots. Usually, the network nodes and their links are set according an application's need, lacking a specific methodology. However, time interval selection is still an open subject. It directly affects the dynamics of simulated epidemics and information spread, mixing properties of random walk, synchronization on networks \cite{Rocha_2017}, and various analysis such as community detection \cite{Medo2018,Narimene2018}, link prediction, attribute prediction, and change-point detection \cite{Fish2017}. Recently, several studies are dedicated to \textit{find proper time intervals} \cite{Zhang2012,Fish2017,Medo2018,Darst2016,Soundarajan2016,Uddin2017,Sulo2011,Narimene2018}. It is widely accepted that the system's time span should be divided into time intervals that are neither small enough to make the network noisy nor large enough to ignore significant time-dependent effects on the network \cite{Fish2015,Soundarajan2016,Sulo2010}. Previous studies report that computation of a proper time interval inline with such limitations is not an independent problem but is defined according to the analysis method used in the application \cite{Sulo2010,Uddin2017,Krings2012,Fish2015,Darst2016, Fish2017, Clauset2012}. A common methodology is generating a time series of topological properties for different window sizes and selecting the window size, which gives the most appropriate time series. 

This study's primary purpose is to propose a generic definition of the proper dynamic network extraction problem to overcome two key concerns reported in previous studies; using network topological properties to measure system stability and applying user-dependent time series analysis methods. This common approach suffers in two ways; first, because most of the topological properties are not only not scale invariant but also not considers all components of the snapshots, and second, they do not allow to make an objective decision change points. We define the generic problem as partitioning the continuous-time span according to time points at which the system's stability is broken. In order to overcome previous issues, we track the system stability by measuring consecutive snapshots' similarity. We develop three scale invariant metrics for all components of snapshots; node, link, and neighborhood similarities. We consult statistical randomness limits of the proposed metrics rather than using a specific analysis method or a predefined threshold. This limit is determined by the null model, which is proposed in \cite{Jaccardlimit}. 
The time intervals whose similarity scores are lower than these limits are the critical time intervals where the system change is more significant than chance. 
We validate the usability of proposed metrics on two different data sets, which have different temporal change trends. The first data set is Wi-Fi access points (WAP) logs of Sabancı University Tuzla Campus, Istanbul. In \cite{Kjaergaard2012}, the authors explain in detail the challenges of using WAP logs for extracting social information and report several problems limiting the handling of raw WAP logs. One needs to model this data accurately to overcome these limitations. The second one is Enron data set \cite{Sulo2010,Fish2017,Darst2016,Clauset2012}. It is a well-known data that is used for a similar purpose as ours. To validate the performance of proposed metrics, we compare them with Clauset's adjacency correlation coefficient \cite{Clauset2012}, which is another metric developed for the same purpose. We concentrate on the noise and informativeness of the signals at the comparison.

The main contributions of this work are three folds. First, we propose an application and method independent definition for a proper dynamic network extraction problem. It is based on measuring systems’ stability objectively. Second, we propose three scale invariant network similarity metrics and use a null model for their statistical significance. Thus, we can identify critical change points objectively without consulting user-defined thresholds or data labeling. Third, we validate similarity metrics on two data sets. We evaluate similarity metrics' performance by comparing them with adjacency correlation coefficient in terms of the noise and diversity of generated signals. The readers find the details of problem definition and proposed methodology in section \ref{Method} and the data set descriptions in section \ref{Datasets}. We give empirical results in section \ref{Results}. Afterward, in section \ref{Conclusion}, we summarize the overall study and discuss its perspectives.

\section{Literature Review}
Finding proper time intervals for dynamic network extraction has mostly been handled with an empirical procedure. Usually, the most proper time interval is selected from a set of predefined candidate window sizes. First, different dynamic networks for each window size is extracted. Second, time series are generated from network snapshots' features for each dynamic network. Third, the robustness of the time series are studied in terms of their noise and informativeness. The two most critical points in these approaches are to extract a time series with appropriate snapshots' features and analyze them with a reliable methodology.

Most of the previous works use topological properties of snapshots as the features for time series generation \cite{Sulo2010,Uddin2017,Soundarajan2016}. However, although the network changes considerably, topological properties' may stay the same or close. 
More importantly, most of them are not scale invariant. Their values depend on the studied snapshot. For this reason, one cannot use them for comparing different snapshots. Although their time series' analysis gives an intuition about the system's proper time intervals, their usage does not constitute a generic approach independent of system properties. The robustness of a generated time series is usually stated by statistical time series analysis \cite{Sulo2010}, or unsupervised \cite{Uddin2017} or supervised learning \cite{Fish2017}. These approaches demand a user-defined threshold, deciding the number of divisions or labelling the data, making them subjective and user-dependent. Moreover, most of those approaches divide the time span into regular time intervals. Only a few works find time intervals with different duration \cite{Soundarajan2016}.

In the literature, we encounter that snapshots' spectral properties \cite{Clauset2012} or similarity metrics \cite{Darst2016} are used for overcoming the limitations of topological properties. Clauset and Eagle \cite{Clauset2012} propose one of the earliest solutions. They made a time series spectral analysis on not only topological properties as degree and transitivity but also on the adjacency correlation coefficient that they developed for measuring nodes' neighborhood similarity between consecutive snapshots. In \cite{Darst2016}, the authors underline the roles of similarities of consecutive system events for the detection of time intervals. They look for the peaks in the Jaccard similarity of the system events between a given time and after a precise time interval based on a similar assumption proposed in this work. Nevertheless, they use system events similarity directly and not the consecutive snapshots' similarity. Krings et al. proposed a similar solution in \cite{Krings2012}. Unlike the approach of \cite{Darst2016}, they do not use system events' similarity. They use both topological properties and Jaccard similarity-based consecutive snapshot similarities. However, when they slide windows, they do not extract separate snapshots but aggregate them. After all the sliding process, they extract a single big complex network that represents all system activities. They examine the results of topological properties and similarity scores at each slide, focusing on only link set similarities.

\section{Proper Dynamic Network Extraction}\label{Method}
\subsection{Problem Definition}
Let us consider $\mathcal{C}$ is a complex system which includes interacting objects. The beginning and end timestamps of the interactions are $t_1$ and $t_{\theta}$ respectively. The interactions change in a \textit{continuous time interval} $\left[t_1,t_{\theta}\right]$. A static network $G=(V,L)$ is a pair such that $V$ is set of nodes and $L\subseteq V \times V$ is its set of undirected links. A \textit{dynamic network} for representing $\mathcal{C}$, $\mathcal{G}=\langle G_1,\ldots,G_\theta \rangle$, is a finite sequence of chronologically ordered static networks $G_i$ ($1\leq i\leq\theta$). We call each of them, $G_i$, snapshots. More formally, a \textit{snapshot} $G_i=(V_i,L_i)$ is a pair of node set and link set where $i$ is a sub-interval in $\left[t_1,t_{\theta}\right]$. Hence, $V_i$ is the set of nodes representing the objects appearing during $i$ in $\mathcal{C}$ and $L_i \subseteq V_i \times V_i$ is the set of their links. \textit{Duration} of the snapshot $G_i$ is the length of sub-interval $i$. 

The \textit{shortest stable duration}, $\epsilon$, of $\mathcal{C}$ is the shortest duration in which the system stays stable. It corresponds to a period where there are few changes in the system objects and interactions from the beginning and end of that period. In order to determine \textit{the longest stable duration}, $\bigtriangleup_i$, for given beginning time $t_i$, we add $\epsilon$ to $t_i$ until the stability of system breaks. Thus, $\bigtriangleup_i$ is a multiple of $\epsilon$. Considering that $\epsilon$ corresponds to a concise moment, the completeness of $\bigtriangleup_i$ is the longest duration is evident. Once the system loses its stability, the rest part could be seen as a new system in order to determine the new longest stable duration, $\bigtriangleup_{i+1}$, of the beginning time $t_i + \bigtriangleup_i $. Hence, $\bigtriangleup_{i+1}$ could be found as another multiple of $\epsilon$ where the system's stability is broken. By this iterative approach, we define \textit{proper duration sequence}, $\bigtriangleup=\langle \bigtriangleup_1,\ldots,\bigtriangleup_{\theta-1} \rangle$ is the sequence of the longest stable duration in which $\mathcal{C}$ is stable. Accordingly, $\langle \left[t_1,t_1+{\bigtriangleup_1}\right],\ldots,\left](t_1+\scriptstyle\sum_{i=1}^{\theta-1}\bigtriangleup_i),(t_1+\scriptstyle\sum_{i=1}^{\theta}\bigtriangleup_i)\right] \rangle$ is \textit{proper discrete sub-interval division} of $\left[t_1,t_\theta\right]$. $\mathcal{C}$ is stable at its longest duration inside the period of any member of this sequence, however between consecutive members, $\mathcal{C}$ is unstable. We define a \textit{proper dynamic network} of $\mathcal{C}$ as $\mathcal{G}_{\bigtriangleup}=\langle G_{\left[t_1,t_1+{\bigtriangleup_1}\right]},\ldots,G_{\left]t_1+{\sum_{i=1}^{\theta-1}\bigtriangleup_i},t_1+{\sum_{i=1}^{\theta}\bigtriangleup_i}\right]} \rangle$. Each member of $\mathcal{G}_{\bigtriangleup} $ is the snapshot for given discrete sub-interval. By this definition, we expect that for any snapshot, $\mathcal{C}$ stays stable at its longest duration till the beginning of next snapshot. \textit{Proper dynamic network extraction} is the problem of finding proper discrete sub-interval division of the time span where the system is defined. It can also be seen as a continuous time discretization by considering the stability of system. 

\subsection{Measuring the Stability }\label{toposim}

Let $S$ be a metric for quantifying the \textit{stability} of unchanging components, i.e., objects and interactions, at given sub-interval. If $S$ is large, $\mathcal{C}$ is stable from the beginning till the end of the a given sub-interval. To extract a proper dynamic network from $\mathcal{C}$, one needs to shift a window of $\epsilon$ length starting from the first time point in the direction of time flow until the result of $S$ is showing that stability is broken. Then the same window shifting process continues recursively for the unprocessed time until the whole system is scanned. There should be an objective bound of $S$ to decide whether the system's stability is broken.
$S$ can be any network similarity measure whose objective bound is defined. 
In the literature, many similarity measures of complex networks were discussed in \cite{Schieber2017,Wills2019}. They focused on networks' spectral properties; nevertheless, objective bounds of those measures are not defined. We propose to use \textit{Jaccard Similarity} which quantifies the number of common parts of two different sets \cite{Jaccard12similarityCoefficient}. Its statistically expected value is also defined under a null model \cite{Jaccardlimit}. 

We use Jaccard Similarity on three different types of sets. For the sake of comprehensibility, we use different similarity names dedicated to different types of sets. 
The first and second similarities are \textit{Node Similarity}, $S_{node}(t\_prev,t\_next)$ (equation \ref{NodeSim}) and \textit{Link Similarity}, $S_{link}(t\_prev,t\_next)$ (equation \ref{LinkSim}). They are Jaccard Similarities of nodes sets and links sets of previous and next snapshots respectively. 

\begin{equation}\label{NodeSim}
S_{node}(t\_prev,t\_next)=\frac{ \mid{V_{t\_prev} \bigcap V_{t\_next}}\mid}{\mid{V_{t\_prev} \bigcup V_{t\_next} }\mid}
\end{equation}

\begin{equation}\label{LinkSim}
S_{link}(t\_prev,t\_next)=\frac{ \mid{L_{t\_prev} \bigcap L_{t\_next}}\mid}{\mid{L_{t\_prev} \bigcup L_{t\_next} }\mid}
\end{equation}

Let $N_{t\_prev}(v)$ and $N_{t\_next}(v)$ are the first order neighborhood of $v \in V_{t\_prev} \bigcap V_{t\_next}$ at previous and next snapshots respectively. We define for a given node, $v$, its \textit{neighbor stability}, $\delta(v,t\_prev,t\_next)$ as Jaccard Similarity of $N_{t\_prev}(v)$ and $N_{t\_next}(v)$ (equation \ref{NodeNeigborSim}). 

\begin{equation}\label{NodeNeigborSim}
\delta(v,t\_prev,t\_next){\frac{ \mid{N_{t\_prev}(v)\bigcap N_{t\_next}(v)}\mid}{\mid{N_{t\_prev}(v) \bigcup N_{t\_next}(v) }\mid}}
\end{equation}

The third similarity is the \textit{Neighborhood Similarity}, $S_{neighbor}(t\_prev,t\_next)$ (equation \ref{CommonNodeSim}). It is the average neighbor stability of the common nodes in previous and next snapshots. $S_{neighbor}$ uses nodes as units. It reflects the average neighborhood stability over nodes while $S_{link}$ concentrates directly on links. If network node numbers do not change too much, and network centralization is not large, $S_{link}$ and $S_{neighbor}$ contains similar information. However, if there are considerable nodal changes, the result scores of those two metrics fall apart. For example, for a star-shaped network whose most of its non-central nodes leave as time goes by, it's $S_{link}$ and $S_{neighbor}$ would be different.

\begin{equation}\label{CommonNodeSim}
S_{neighbor}(t\_prev,t\_next)=\frac{1}{\mid{V_{t\_prev} \bigcap V_{t\_next}}\mid}\sum_{v}\delta(v,t\_prev,t\_next)
\end{equation}

In \cite{Jaccardlimit}, the authors study the limits of Jaccard Similarity with the probabilistic approach for the biological similarity of species for the taxonomy. They propose a null model (equation \ref{JaccardBound}) that quantifies how much random the two studied sets are in terms of their common element numbers. They also give the list of critical values of Jaccard Similarity in terms of a total number of elements for small sized sets in \cite{Jaccardlimit2}. They work with the samples of species. When two species are compared, the total possible number of attributes that can be observed is noted as $N$. $A$ and $B$ are the numbers of attributes present in samples of first and second species, respectively, and $C$ is the number of attributes present in both species in the samples.

\begin{equation}\label{JaccardBound}
P=\frac{\sum\limits_{x=0}^C {A+B-x \choose x}}{\sum\limits_{x=0}^{min(A,B)} {A+B-x \choose x}}
\end{equation}

We can interpret $A$ and $B$ as the number of elements in two sets. The common elements may range from $0$ to the minimum value of $A$ and $B$. Thus, the denominator of equation \ref{JaccardBound} represents the number of all possible intersection states that two sets can have. Similarly, the nominator expresses the number of all possible intersection states that two sets can have when knowing that two sets have at most $C$ elements in common. Finally, the equation \ref{JaccardBound} gives the limit of having $C$ elements in common by chance. If two sets are similar, their Jaccard Similarity should exceed the limit of being by chance. In our problem, we will compare two consecutive snapshots' node sets or link sets. Because node numbers or link numbers are larger than the table given in \cite{Jaccardlimit2}, we apply the formula given in equation \ref{JaccardBound}. In our case, $A$ and $B$ correspond to the number of nodes or links in consecutive snapshots, and $C$ corresponds to the number of common elements in node sets or link sets for consecutive snapshots. For now, we do not consider $S_{neighbor}$ because $S_{neighbor}$ is an average score calculated for every node. The formula given in equation \ref{JaccardBound} is suitable for computing a limit for $\delta$ for every node, but one cannot use it for $S_{neighbor}$. 

\subsection{Choosing $\epsilon$}\label{setup}
\begin{figure}[!htb]
 \centering
 \includegraphics[scale=0.45]{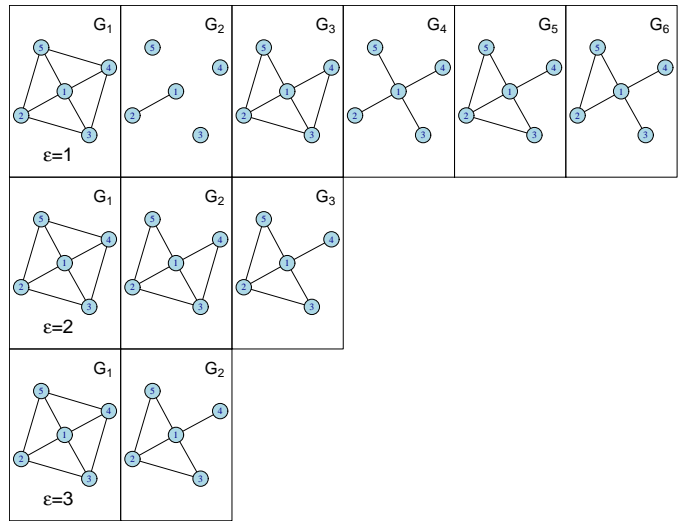}\
 \caption{An example of consecutive snapshots extracted for different $\epsilon$. The top one is extracted for the shortest $\epsilon$, one unit. The network has six snapshots. Consecutive ones seem dissimilar with each other. The center one is extracted for $\epsilon$ two units. Three snapshots are looking like each other. The bottom one is extracted for $\epsilon$ three units. The network has two snapshots again looking like each other. We cannot catch the dynamic changes at this one as well.}
 \label{fig:epsilon_importance}
\end{figure}

The choice of $\epsilon$ has a critical effect on determining a system's stability. Figure \ref{fig:epsilon_importance} reveals how the choice of $\epsilon$ can change the resulting network structure on three dynamic networks for the same system. In this example, the network is changing fast. Thus network dynamism is affected in a negative way when we increase $\epsilon$. In another example, if the system changes slowly or stays stable for a too long time, large $\epsilon$ might bring advantages.

The proper value of $\epsilon$ is between two limits; either small, making the network too noisy, or large, making it redundant and not informative. In \cite{Sulo2010}, the noise and the information of time series are measured by variance and compression ratio, respectively. The proper $\epsilon$ corresponds to the difference between variance and compression ratio is lower than a user-defined threshold. We also use similar statistics for comparative analysis. Given a fixed $\epsilon$ and the corresponding $\mathcal{G_{\epsilon}}$, $\mathcal{F_{\epsilon}}=\langle F_1,\ldots,F_t\rangle$ is time series of a similarity score. We measure the noise of $\mathcal{F_{\epsilon}}$ by its variance and normalized standard deviation. Their formula is given in equations \ref{variance} and \ref{normalized_sd} respectively.

\begin{equation}\label{variance}
\sigma^2 = \frac{1}{t}\sum_i^t(F_i-\mu(\mathcal{F_{\epsilon}}))^2
\end{equation}

When $\epsilon$ is too small, the noise of the time series is too low because each consecutive snapshots look like the other. As a result, they have large and same similarity scores. Contrarily, when $\epsilon$ is too large, noise is too low again. All snapshots are different. It results in low and close similarity scores for all snapshots, with a low variance of time series. 

\begin{equation}\label{normalized_sd}
\sigma_n = \frac{\sigma}{\mu(\mathcal{F_{\epsilon}})}
\end{equation}

Large values of variance indicate $\mathcal{F_{\epsilon}}$ changes drastically in time, making it hard to distinguish between the occurrence of a meaningful change and a noise effect. On the other hand, small values of variance indicate $\mathcal{F_{\epsilon}}$ is smooth, and a lot of the noise is removed. For a proper $\epsilon$, the time series variance should not be large enough to be considered noisy but also low enough not to become informative. Variance is not scale invariant. Its use for comparison may be incomplete. That is why we also consult the normalized standard deviation given in equation \ref{normalized_sd}. This normalization allows the spread of a variable's distribution with a large mean and corresponding large standard deviation to be compared more conveniently with the spread of the distribution of another variable with a lower mean and a correspondingly lower standard deviation. Normalized standard deviation is independent of residual units and more convenient for comparative studies. We also use two statistics for stating the diversity of time series. Their general formula is given in equation \ref{smoothness}.

\begin{equation}\label{smoothness}
s = \frac{c}{u}
\end{equation}

Here $u$ is the length of the string representation of $\mathcal{F_{\epsilon}}$, and $c$ is the length of its compressed representation. We consult two compression techniques. The first one is \textit{String Compression} by run-length encoding. This takes into account the same consecutive values when compression is performed. The latter one is basic compression. It deals with unique values without considering sequential sameness. A small value of $s$ represents a high compression state. It means there are many redundancies in the signal. In other words, the signal has less noise and is smooth. A good $\epsilon$ corresponds to the value where $s$ is large with variance is low. We call \textit{string diversity} and \textit{non-repetition level} for these two statistics for the rest of the article.

We compare the proposed similarity metrics performance with \textit{adjacency correlation coefficient}, $\gamma$. $\gamma$ is also developed for measuring the similarity of consecutive snapshots by \cite{Clauset2012} for adjusting $\epsilon$. It is the correlation of adjacency matrices of consecutive snapshots. Comparative performance analysis of similarity metrics with $\gamma$ by their sensitiveness to reflect the noise and diversity level for different $\epsilon$ is explained in section \ref{Results}.

\section{Data Sets}\label{Datasets}
In our experiments, we use both the Sabancı WAP log and the Enron email data set. Because the Sabancı WAP log has not been analyzed as dynamic networks, we also examine its dynamic networks' topological properties. Enron email set is used for the same purpose of ours by \cite{Sulo2010,Fish2017,Darst2016,Clauset2012} before. 

\subsection{Wi-Fi Access Point Connections}
\begin{figure}[!ht]
	\centering
	\includegraphics[width=6cm]{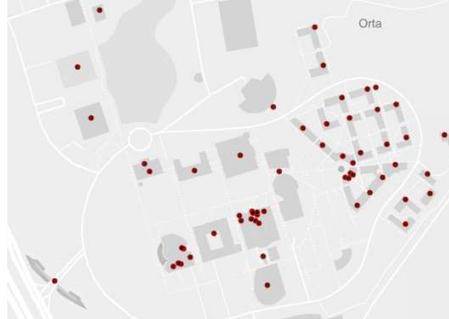}
	\caption {Spatial Wi-Fi access point distribution in the campus.}
	\label{fig:campus}
\end{figure}

The Wi-Fi tracking data consists of system connection metadata, covering the 2016-2017 fall semester, 137 days in total, that were logged in every $10$ minute by the IT department of Sabancı University. At each logging moment for each access point, device IDs connected to that access point are received with a time stamp. The campus consists of 3 major areas: dormitories, faculty buildings, and utility centers. The lessons start at 8:40 and end at 19:30. Each lesson is $50$ minutes, and there is a $10$ minutes break between them. The log records, 9 million in total, include \textit{device ID, connection/disconnection timestamps} and \textit{WAP name}. In total, the device IDs, 68.114, were anonymized by assigning unique values to each MAC address that connects to any WAPs on the campus. We create a node for each device ID that appears in the system. If two devices connect to the same Access Point for a given time interval, we put a link between them. A link between two nodes might be a sign that those devices are at the same place. The majority of the WAPs are located in dormitories and faculty buildings, while utility centers possess a small portion. The distribution of 613 WAPs can be observed in figure \ref{fig:campus}.
Collected data from WAPs have much noise and require cleaning. First, we remove the records whose connection and disconnection timestamps are not represented. Then, we merge the records representing the connection's continuity but dropped because of WAPs' properties. Once a connection drops, a device can connect to the same WAP or another close to the previous WAP. In this work, we consider the first case. For each device, we order the data according to their connection time. If two consecutive records have the same WAP names and the previous records' disconnection time is equal to the next records' connection time, we merge these two records. We use eight different values in minutes, $\epsilon = \left\lbrace 1, 3, 5, 10,30, 40,60 ,1440 \right\rbrace$ for candidate dynamic networks extraction. We choose these values to respect the minimum ($1$ min.) and the most frequently seen ($3$ and $5$ min.) periods, break times between lectures ($10$ min.), proper walking duration in the campus ($30$ mins.), proper lecture duration ($40$ and $60$ min.) and one day ($1440$ min.). 

\subsection{Topological Properties of WAP Snapshots}\label{sec:topologyWAP}
\begin{figure}[!htb]
 \centering
 \includegraphics[scale=0.8]{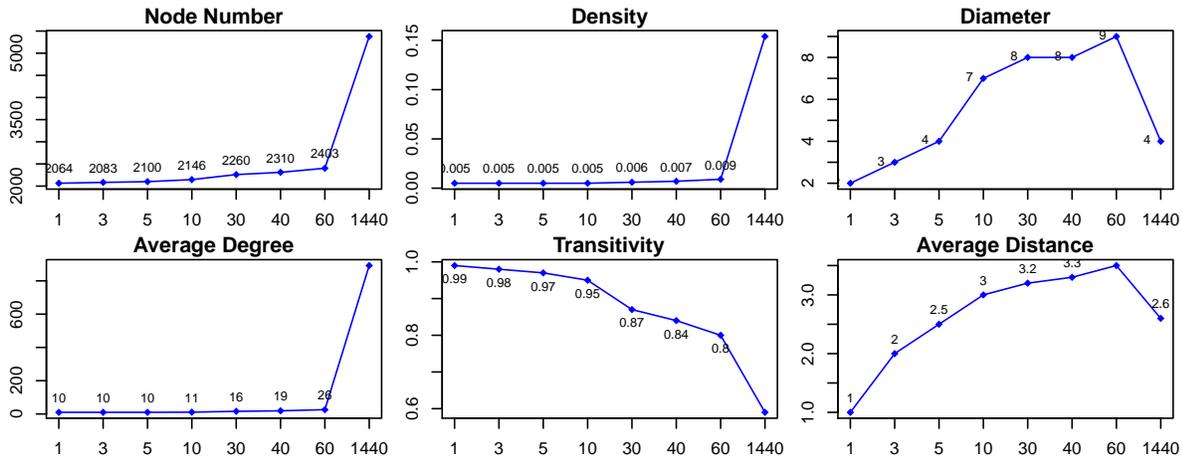}
 \caption{Average values of topological properties. The x-axis represents the corresponding $\epsilon$ and the y-axis represents the average value of the corresponding topological property.}
 \label{fig:Topologic_average}
\end{figure}

We examine node and link number, density, average degree, number of the connected component, average path length, diameter, transitivity and closeness, and betweenness centralities. We show average values of topological properties in function of $\epsilon$ in figure \ref{fig:Topologic_average}. Snapshots have similar topological properties with well-known social and information networks \cite{newman2003structure}. As $\epsilon$ increases, on average, we have more crowded, denser, and more connected snapshots with a shorter average distance. The most striking point is that when $\epsilon$ is $1440$, the snapshots have, on average, quite a different topology than those produced for other $\epsilon$. The average density is too large ($\sim 0.15$) with a large average degree ($892$) and relatively lower transitivity ($\sim 0.59$). Those snapshots do not exhibit realistic behavior. It seems this $\epsilon$ is too large for this system. Thus, we do not examine their results for further analysis. 

\begin{figure}[!htb]
 \centering
 \includegraphics[scale=0.7]{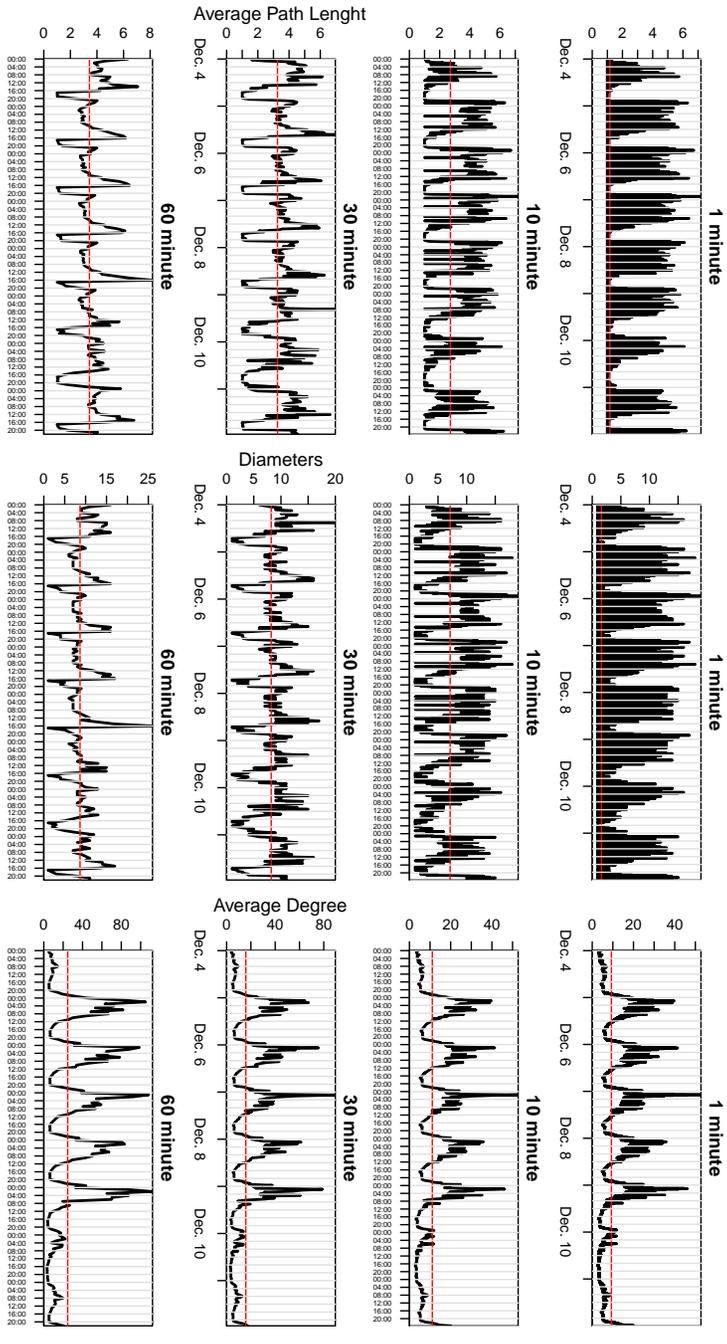}
 \caption{Signals for the topological properties of the different dynamic networks extracted for WAP logs in a week between 4-12 December 2016. Each plot shows the results for different $\epsilon$. The X-axis shows both date and time information. Y-axis shows the average path length, diameter, and average degree for the left, middle, and right blocks, respectively. A red line on each plot shows the average value of the related signal.}
 \label{fig:Topologic2}
\end{figure}

Time series of each topological properties has stationary signals with the daily cycle on weekdays. The noise level depends on the studied $\epsilon$. We show average path length, diameter, and average degree signals in figure \ref{fig:Topologic2}. The signal routine seems to be broken at weekends, corresponding between 10 and 12 December. Topological properties' values are lower than the general range at these days. Moreover, signal trends are different from other days. Considering all topological properties, in general three critical time intervals that the daily cycle occurs are (1) between $[00:00-02:00]$ and $[08:00-10:00]$, (2) between $[08:00]$ and $[17:00-20:00]$ and (3) between $[19:00-21:00]$ and $[00:00-02:00]$. Those time intervals correspond to three major periods of a day; \textit{night}, \textit{day time} and \textit{evening to night} respectively. Ascending and descending parts of the signals correspond to the hours when the campus gets active and calm respectively.

Circadian rhythms of human activity can be extracted from electronic records \cite{Talayeh_circadian, song2010limits,Jo_2012}. In \cite{Talayeh_circadian}, the authors underline two major periods of circadian rhythm; \textit{day time} and \textit{night}. They reveal that different social profiles exhibit different behavioral patterns at those periods. Some people are active at night while some others are not. Our findings are also supporting them partially. Looking at details inside circadian rhythms of campus, we distinguish different social groups as well. For instance, at \textit{night} period, campus seems too active. People are getting into campus or walk around, connecting to different WAP's. Most of the night activities are due to WAP connections in the dormitory area. In the \textit{day time}, the campus is again active and crowded, but this time, not only in the dormitories but also at faculty buildings and utility centers. At \textit{evening to night} period, the campus seems calm with few activities. This period corresponds to the hours when the lessons have finished, employees and faculty members leave the campus, and students go back to the dormitories. We also remind that weekend activity is low. Interpreting all these findings, we distinguish two major groups on the campus; \textit{people living on campus and off campus}. Most of the people from the first group also leave the campus during the weekend, which results in low activity. 

\subsection{Enron Emails}
Enron is a well-known dataset that has been widely studied in network science. This study generates dynamic networks by dividing raw Enron data into different time intervals to find the most proper time interval for its modelling. Raw data consists of the emails of 151 Enron company employees. These emails are sent between 1997 and 2002. We consider company employees and all mail addresses at every From and To fields in the corresponding emails as network nodes. As a result, 28.802 nodes are generated. A link is established between two nodes if they sent email to each other for a given time period. Although email systems are usually modelled under directed networks, we generate undirected snapshots because our focus in this study is to determine the proper time intervals. We analyze 1.178 days of data. Snapshots for short $\epsilon$ like a period of minutes or hours were too sparse, and they have non realistic topological properties. That is why; we use longer periods on a daily scale. We choose $\epsilon = \left\lbrace 1, 7, 15, 30, 90,180 \right\rbrace$ days. We respect the daily, weekly, monthly, quarterly, and half-year periods for companies.

\section{Results}\label{Results}
In the following section, we report comparative analysis and expected limits of similarity metrics for WAP and Enron snapshots, respectively.

\subsection{Similarity Results of WAP Snapshots}
\begin{figure}[h]
 \centering
 \includegraphics[scale=0.75]{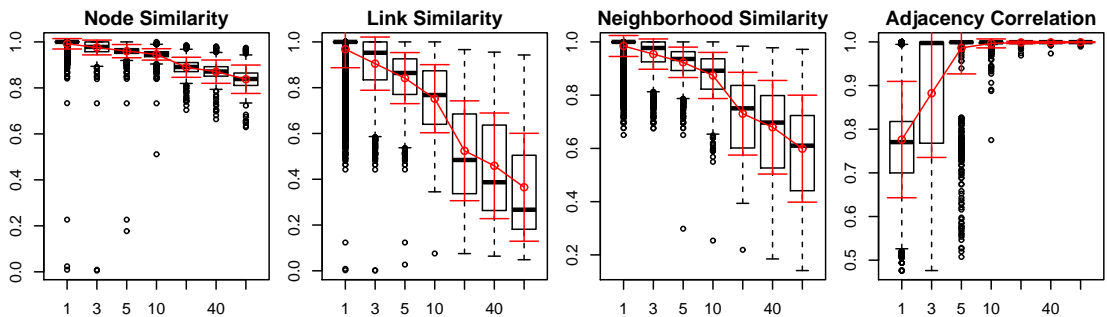}
\caption{Box plots of similarity scores and $\gamma$ for WAP snapshots. X-axis is $\epsilon$. Red lines are the average values of time series while error bars are their standard deviation.}
\label{fig:boxSimilarities}
\end{figure}

We represent the box plots of similarity metrics and $\gamma$ in function of $\epsilon$ in figure \ref{fig:boxSimilarities}. Accordingly, similarity scores' reaction to $\epsilon$ increase looks like each other, while $\gamma$ behaves differently. Proposed similarities decrease as $\epsilon$ increases. However, $\gamma$ first increases for $\epsilon<10$ minutes. Then, it stays stable at its maximum value, $1.00$. It means that $\gamma$ cannot distinguish the effect of $\epsilon>10$. The most and least sensitive metrics to $\epsilon$ changes are $S_{link}$ and $S_{node}$ respectively. The largest and the lowest decreases are seen on them respectively. 
Especially, for $\epsilon=60$, snapshots seem to have different link structure ($\mu(S_{link}) \sim 0.36$) while most of the network nodes stay same ($\mu(S_{node}) \sim 0.84$). Regarding the spread of the values in box plots, $S_{node}$ is noisy on all $\epsilon$. $S_{link}$ and $S_{neighbor}$ are noisy only when $\epsilon<30$. $\gamma$ produces a noisy signal in all but $\epsilon=3$. For $3$, it produces a signal with a score of $1.00$ for the majority of the snapshots. According to this metric, the majority of consecutive networks are similar to each other. It does not differentiate $\epsilon$ effect for larger values of $3$.

\begin{figure}[!htb]
 \centering
	 \includegraphics[scale=0.7]{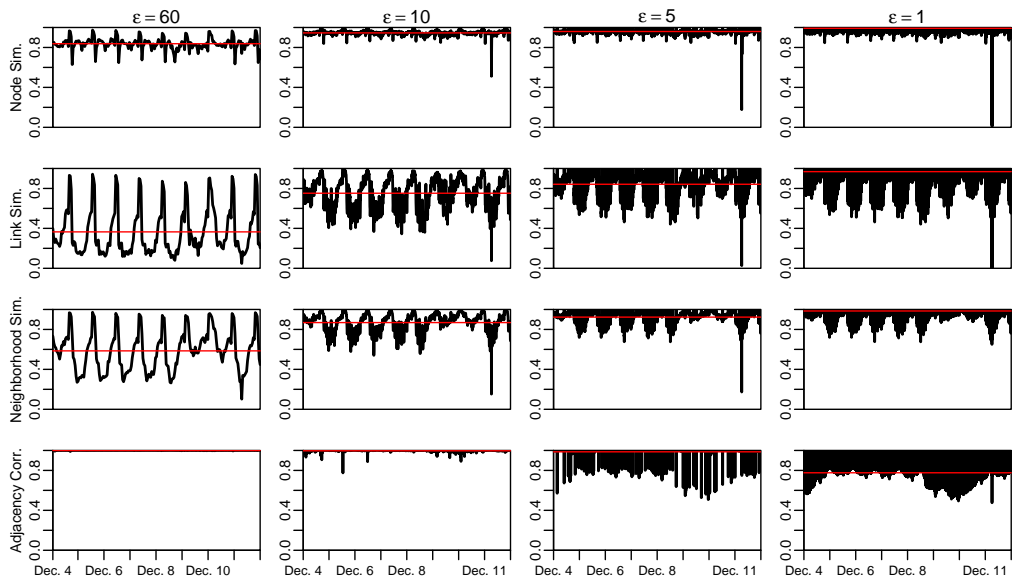}
	\caption {Time Series of similarity metrics and $\gamma$ for the snapshots of $\epsilon = \left\lbrace 1, 5, 10, 60 \right\rbrace$ minutes for WAP snapshots. Red lines are the average values of time series.}
	\label{fig:scores}
\end{figure}

Focusing on each metrics' signal (see figure \ref{fig:scores}) in detail, like topological properties (see section \ref{sec:topologyWAP}), proposed similarities signals are also stationary and have cycles. However, $\gamma$ signal seems different from them. It does not seem to be sensitive to $\epsilon$ change. For $\epsilon<10$, its signal seems too noisy, while it is not informative for larger values. It takes the same values as time goes by and does not reflect the campus's circadian rhythm at its signals. Nevertheless, proposed similarity results support the circadian rhythm explained in section \ref{sec:topologyWAP}. Accordingly, the activity in the campus \textit{increases} from $07:00$ till $18:00$, \textit{decreases} from $18:00$ till $00:00$ and \textit{stays stable} from $00:00$ till $07:00$. These time intervals are compatible with different parts of the day we find in section \ref{sec:topologyWAP}. Therefore, the topological properties gave us more vague intervals. Similarity metrics reveal clearer time intervals.

\begin{figure}[!htb]
	\centering
	\includegraphics[scale=0.60]{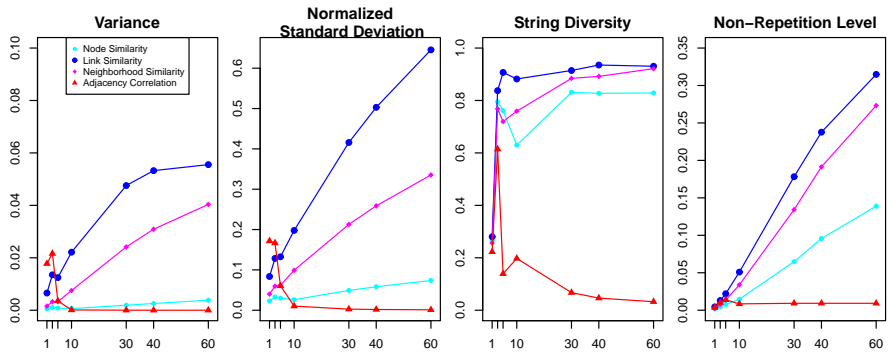}
	\caption {Noise and diversity scores of similarity metrics and adjacency correlation coefficient with respect to $\epsilon$ for WAP snapshots.}
	\label{fig:variationAndStability}
\end{figure}

We compare the performance of the proposed similarities with $\gamma$ by the noise and information levels of their time series. In figure \ref{fig:variationAndStability}, variance, normalized standard deviation, string diversity and non-repetition level in function of $\epsilon$ of all 4 metric signals are given. The larger the $\epsilon$, the more noisy the signals of the proposed three metrics but the smoother the signals of $\gamma$. Evaluating proposed three metrics among themselves, $S_{node}$ is not sensitive to $\epsilon$ increase. $S_{link}$ is the most sensitive one. $S_{neighbor}$ is in-between. $S_{link}$ shows a larger variance over a wider range of values for $\epsilon >5$. $S_{neighbor}$ also shows similar behavior with $S_{link}$. However, it is relatively less sensitive for detecting noise. $\gamma$ shows the opposite behavior; up to $\epsilon=5$, the noise level drops and then turns into a completely flat signal. The flat signal is not informative for proper $\epsilon$ selection. The most suitable $\epsilon$ is the one that variances are low, and diversity is large. Accordingly, the most useful metrics seem to be $S_{link}$ and $S_{neighbor}$. $S_{node}$ is not distinguishing too much the factor of different $\epsilon$ and $\gamma$ generates too smooth and not informative signals. 

As the proposed similarities reflect the network topology, one can track the life cycle of the studied system through them. Moreover, $S_{link}$ and $S_{neighbor}$ generate optimal signals for determining proper $\epsilon$. However, $\gamma$ shows neither the system's circadian rhythm nor the effect of $\epsilon$. That is why; it is not suitable for finding a proper time interval.

\subsection{Proper Time Interval for WAP Snapshots}
\begin{figure}[h]
	\centering
	\includegraphics[scale=0.6]{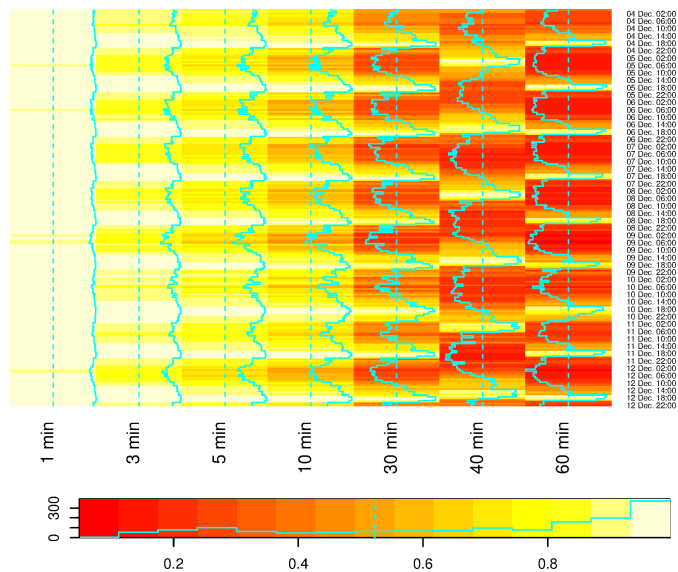}
	\caption {Heat map of the time series, generated for different $\epsilon$, of $S_{link}$ for one week of WAP snapshots. The length of the time series for different $\epsilon$ are fixed to the one of $\epsilon=60$. For the $\epsilon < 60$, the time series are shrunken by averaging the scores. The x-axis represents the corresponding $\epsilon$, and the y-axis represents $S_{link}$ scores in color from red to yellow. The lighter the color, the closer the $S_{link}$ to $1.00$.}
	\label{fig:heatMap}
\end{figure}

Since the metric where we see the effect of $\epsilon$ most clearly is $S_{link}$, we interpret it through its heat map shown in figure \ref{fig:heatMap}. As expected, the larger the $\epsilon$, the lower the $S_{link}$. It also starts to discriminate different periods of the day more significantly. $S_{link}$ shows more considerable variations over a broader range than other metrics. In general, $\epsilon <10$ causes noisy signals for all similarities. These time intervals are too short to distinguish the general differences of snapshots. Short periods can be suitable intervals to follow up on any specific event. However, to understand public life on the campus, they generate noisy results. $60$ minutes look like a reasonable duration for understanding the system's daily routine and clearness of the signal. 
$S_{link}$ being larger on weekends reveals that the campus is more stable and calm over weekends. We distinguish the range of the signals on those days are tight. Moreover, the similarities are more extensive than other days, especially when $\epsilon$ is between $10$ and $60$, as fewer people stay on the campus, and they mostly stay in the dormitories.

\begin{figure}[h]
	\centering
	\includegraphics[scale=0.90]{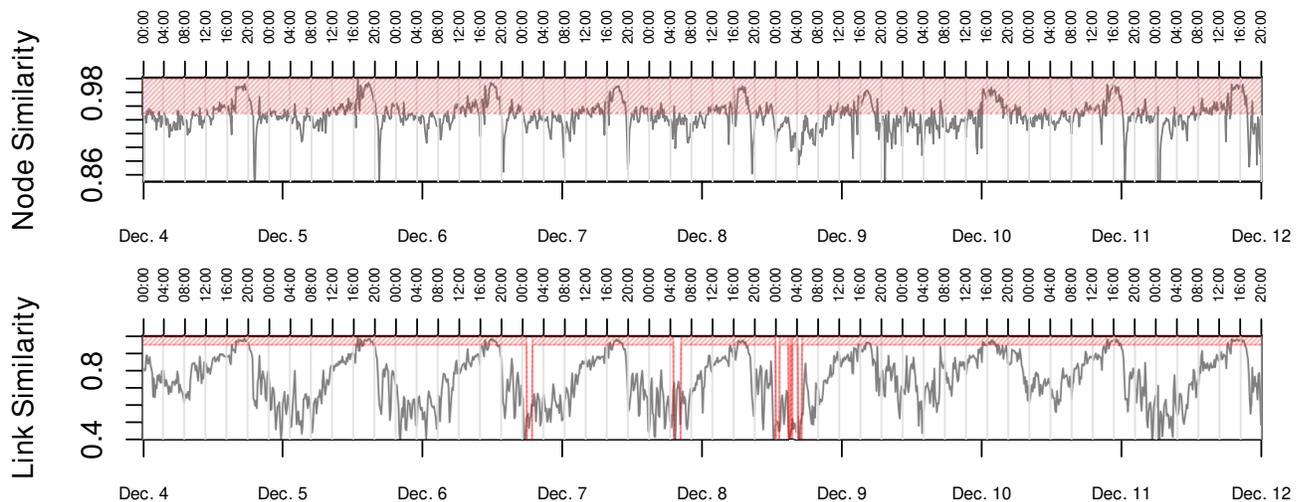}
	\caption {Node and Link Similarity results for $\epsilon =10$ minutes for WAP snapshots. The statistically expected value for each consecutive snapshots is calculated by using equation \ref{JaccardBound}. We also consider the $0.05$ error rate for each expected value. They are indicated as the red hatched zone in the plots. The similarity signal parts that coincide with or above the red hatched zone are the time intervals that the system stays stable in terms of the studied metric. It means that no need to create a new snapshot, but an aggregation of snapshots is possible. The signal parts below the red hatched zones are critical change periods of the system.}
	\label{fig:10_min_limits}
\end{figure}

We consider the expected values of the proposed similarities for finding proper time intervals. Signal meets with or stays outside of their expected value partially for all $\epsilon$. For $\epsilon > 10$, for some time intervals the expected value is $1.0 (-0.05)$. This means that almost no link change in the system is a sign of system stability. For other time intervals, it is $0.00$. Thus, the system is considered stable, even if the link changes at a large rate. When the expected value is close to $0.00$, it seems like the system's continuity in terms of its link structure is lost. The details of expected limit values can be seen in figure \ref{fig:10_min_limits}. Accordingly, $S_{node}$ stays out of the red hatched zone from 02:00 at night till 13:00 in the afternoon and around 21:00 in the evening. In other words, system actors change considerably. Thus, new snapshots can be extracted. The other time intervals look like calm periods of the campus with fewer actor changes. Hence, snapshots can be aggregated at them. Considering $S_{link}$, for everyday, only from 16:00 till 20:00, the scores are larger than their expected value. Thus, the snapshots can be aggregated at this period. Briefly, snapshots can be extracted at ten-minute intervals from midnight to 13:00 during the day while creating the WAP dynamic network. However, snapshots can be extracted for periods of one hour or longer between 13:00 and 20:00 during the day. Extracting snapshots with the same $\epsilon$ for all times will result in redundant snapshots for this system.

\subsection{Enron Results}
\begin{figure}[h]
 \centering
 \includegraphics[scale=0.7]{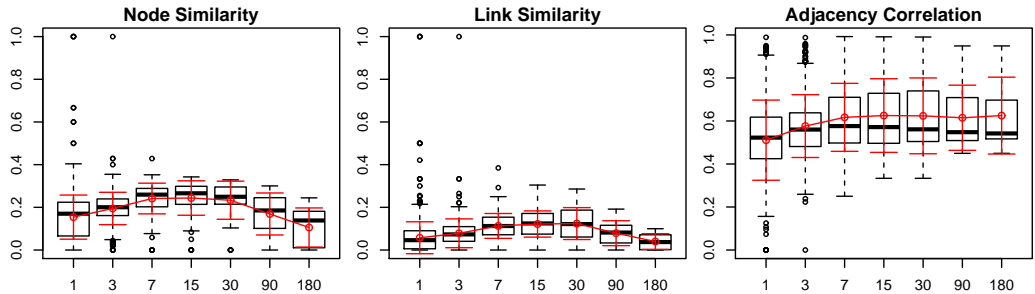}
\caption{Box plots of similarity scores and $\gamma$ for Enron snapshots. X-axis is $\epsilon$. Red lines are the average values of time series while error bars are their standard deviation.}
\label{fig:enron_averageSimilarities}
\end{figure}

Box plots of similarity metrics and $\gamma$ are shown in figure \ref{fig:enron_averageSimilarities}. The signals of these metrics have different characteristics from the ones of WAP snapshots. The main reasons for these differences are the data characteristics and $\epsilon$ time scale. We work with daily intervals in Enron while it was minutely in WAP. In general, the result scores are too low compared to WAP snapshots. Regarding the signal trends, $S_{link}$ and $S_{node}$ shows a signal like an arc. They first increase in all $\epsilon$, then remain constant for a long time and then decrease. This behavior becomes the clearest at $\epsilon = 90$. The node and link structure of snapshots change completely at these intervals. These are the periods when the network gets denser to build a stable structure. Average similarity for both proposed metrics and $\gamma$ increases until $\epsilon$ becomes $15$ days then declines too slowly. $S_{node}$ and $S_{link}$ generate noisy signals until $\epsilon = 15$ while it is $\epsilon=3$ for $\gamma$. 

\begin{figure}[h]
 \centering
 \includegraphics[scale=0.6]{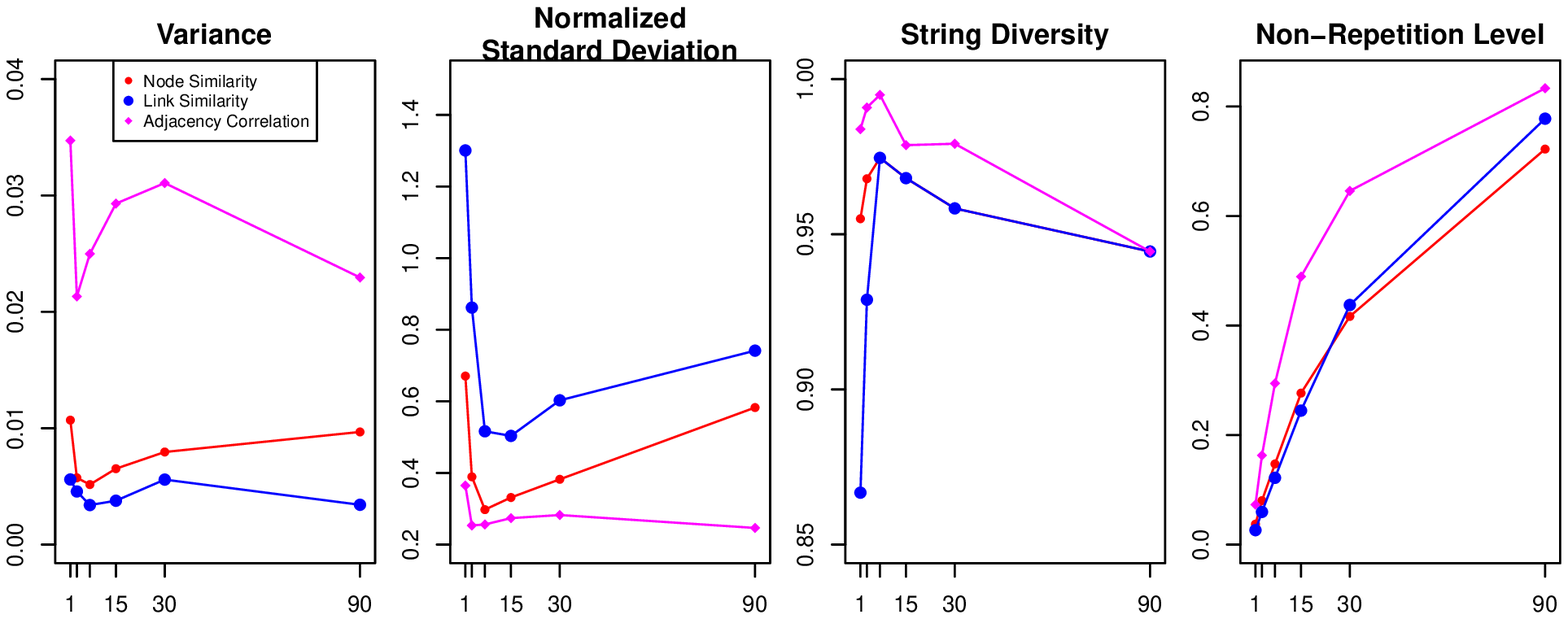}
	\caption{Noise and diversity scores of similarity metrics and adjacency correlation coefficient with respect to $\epsilon$ for Enron snapshots.}
	\label{fig:variationAndStabEnron}
\end{figure}

 \begin{figure}[h]
	\centering
	\includegraphics[scale=0.6]{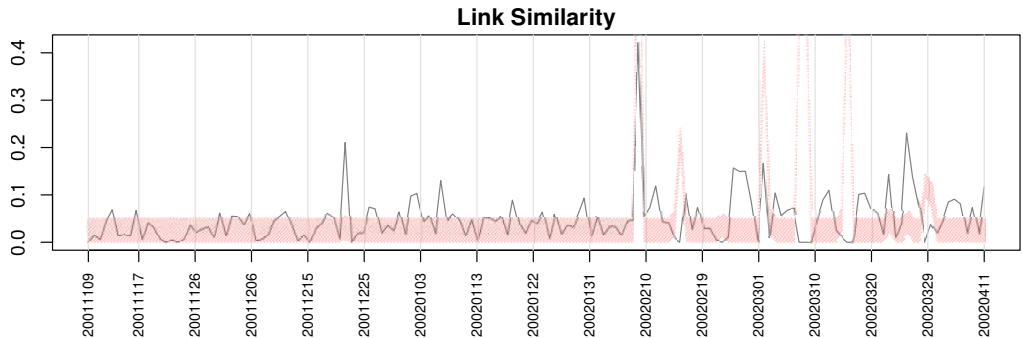}
	\caption {Link Similarity results for $\epsilon =1$ days. The red hatched field is the score that is calculated for the null model. Although the signal fluctuates continuously in the entire time span, it is seen that the change in some days after February 2002, the date that Enron announced its bankruptcy, is more than expected; that is, the system has become dynamic. For days when the signals are below the red hatched area, the network can be cut. Other parts can be aggregated.}
	\label{fig:link_details}
\end{figure}
The comparison of similarity metrics and $\gamma$ in terms of the noise and diversity of the generated signals for different $\epsilon$ is given in figure \ref{fig:variationAndStabEnron}. According to variance, $\gamma$ generates the noisiest signal, while according to normalized standard deviation, it is the least noisy one. This difference is due to the overall scores of $\gamma$ is being large. Normalized standard deviation decreases until an $\epsilon$ value depending on the metric, increasing slightly later. Contrarily, string diversity first increases then decreases. 
$S_{link}$ generates signals with the most different noise and diversity scores for different $\epsilon$ values. In other words, $S_{link}$ reveals the epsilon effect more than other metrics. Different from WAP results, the findings for $\gamma$ signals are similar to the proposed metrics in Enron. In WAP, $\epsilon$ values were short (minutely periods) resulting in a high degree of similarity between consecutive snapshots. $\gamma$ was not sensitive to catch the differences when the similarity was high. However, $\epsilon$ scales are much larger in Enron. This causes the similarity of consecutive networks to be small. $\gamma$ could reflect the high differences like proposed similarity metrics.

Regarding finding the proper time interval for Enron, when $\epsilon=90$, for all snapshots, even if the similarities are too low, it is larger than the expected value. Thus, this time span is too large to capture the significant change. When $\epsilon>1$, the similarity is lower than their expected values in the unstable parts of the system, i.e., at the beginning and end of the time span. That is, the system shows a significant change. When $\epsilon=1$, both $S_{link}$ and $S_{node}$ results are lower than the expected values for many days. Thus, the system shows a significant change. Figure \ref{fig:link_details} reveals the $S_{link}$ trends for the period just before and after Enron announced its bankruptcy.

\section{Discussion and Conclusion}\label{Conclusion}
We propose a formal definition of a dynamic network extraction problem independent from the data or analysis method. This problem can be seen as a compact and informative discretion of continuous time interval where the system is defined. A new snapshot should be extracted if only if there is a considerable change in the system. We propose three network similarity metrics, which are based on well-known Jaccard Similarity, for tracking systems' stability. They are both scale invariant and having statistically expected values. We propose using a null model of those similarities. The values calculated for the null model give us a method independent objective criterion for catching considerable change periods in the system. We compare the effectiveness of proposed similarities with adjacency correlation on distinguishing different time windows for snapshots extraction. We did experiments on two different systems that have different temporal dynamics. The most noticeable results from both experiments can be listed as follows; First, the proposed similarity metrics represent similar signals with network topological properties. Reminding that proposed similarities are scale invariant, they can also be used for any comparison. That is why; they can be used instead of topological properties as they reflect similar information with more objective scoring and more precise signals. Second, in terms of distinguishing the time window for snapshot extraction, proposed similarities generate better signals with more optimal noise and diversity levels than the $\gamma$. $\gamma$ can capture consecutive snapshots differences only when the time window is large. In other words, it is not responsive to the small differences occurring in shorter periods. However, proposed similarities can measure the system’s temporal dynamism more effectively, regardless of the studied data and time window scale. Third, the proposed metrics' statistically expected values can determine the cut-off time points over the entire time span. Using user-defined thresholds in order to decide the significance of similarity metrics are leading to system-specific features to be ignored. In our results, we see that the scores that could be considered low for Jaccard similarity can be larger than their expected values. It allows us to make a more objective evaluation, which is independent of the studied data set. Moreover, we can find out different window sizes for the entire system. In this way, a more compact dynamic representation can be extracted without redundant snapshots. 
\begin{figure}[h]
	\centering
	\includegraphics[scale=0.5]{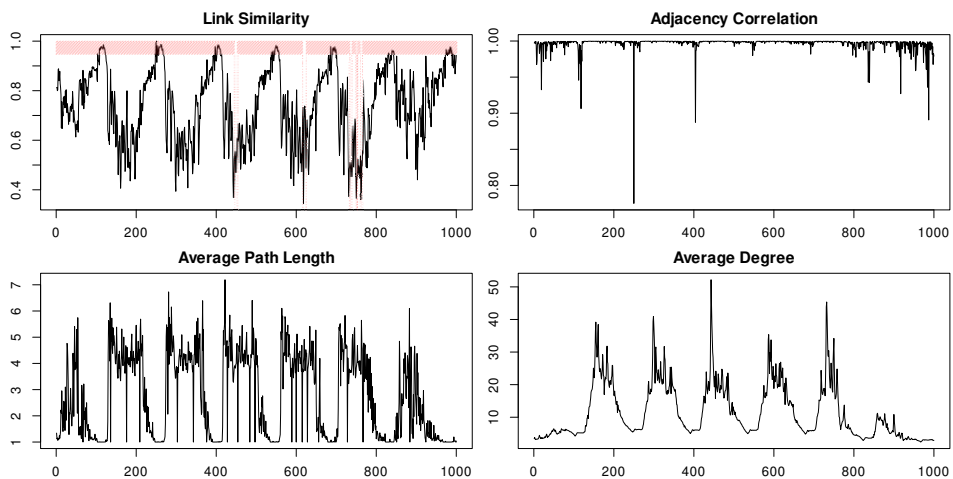}
	\caption {Time series of link similarity, adjacency correlation, and two topological properties for WAP for $\epsilon=10$ minutes. Link similarity shows the circadian rhythm of the system. It has similar trends with topological properties with less noise. Its scores are scale invariant in range $[0.0;1.0]$. Adjacency correlation signals are different from others. They are not informative about campus life. Statistically expected link similarity values allow shifting different sized time windows for the entire network, which in turn does not extract redundant snapshots.}
	\label{fig:link_details}
\end{figure}
 The main limitation of this work is that the proposal focuses on $\epsilon$ choice and the sequential flow in the system. If the system is changing very slowly and chosen $\epsilon$ is too low (or vice versa), consecutive comparison can be meaningless. In this case, there is no large difference between consecutive snapshots, but as time goes by, the further snapshots apart would be quite different. A more self-working system can be built to automate decision-making and overcome the mentioned problem. For example, an iterative method that automatically shifts the time window and aggregates the snapshots whose similarities are higher than their expected values can be an extension of this work. Thereby slowly changing systems' time span can also be cut off correctly because the comparison would be made on a previously aggregated snapshot with a new $\epsilon$ sized snapshot. The work presented here can also be extended to different paths: i- We can enrich the comparison by using different metrics such as adjacency spectral distance. ii-Generated signals can be analyzed by signal processing or regression analysis for stating their noise and diversity levels.

\section*{Acknowledgment}
This article is partially supported by Galatasaray University Research Fund (BAP) within the scope of project number 18.401.004, and titled "Sıralı Sistemlerde Tahmin ve Çıkarım". 

\section*{References}

\bibliographystyle{plain} 
\bibliography{IOPLaTeXGuidelines}
\end{document}